%% file: multiearly_arxiv.tex
\documentclass[article, nojss]{jss}

\author{
Ulrich Beyer \\ F. Hoffmann-La Roche AG, Basel \And 
David Dejardin \\ F. Hoffmann-La Roche AG, Basel \AND
Matthias Meller \\ F. Hoffmann-La Roche AG, Basel \And
Kaspar Rufibach \\ F. Hoffmann-La Roche AG, Basel \AND
Hans Ulrich Burger \\ F. Hoffmann-La Roche AG, Basel
}
\title{A multistate model for early decision making in oncology}

\Plainauthor{Ulrich Beyer, David Dejardin, Matthias Meller, Kaspar Rufibach, Hans Ulrich Burger} 
\Shorttitle{Multistate models for early decision making}

\Abstract{The development of oncology drugs progresses through multiple phases, where after each phase a decision is made about whether to move a molecule forward. Early phase efficacy decisions are often made on the basis of single arm studies based on RECIST tumor response as endpoint. This decision rules are implicitly assuming some form of surrogacy between tumor response and long-term endpoints like progression-free survival (PFS) or overall survival (OS). The surrogacy is most often assessed as weak, but sufficient to allow a rapid decision making as early phase studies lack the survival follow up and number of patients to properly assess PFS or OS.
With the emergence of therapies with new mechanisms of action, for which the link between RECIST tumor response and long-term endpoints is either not accessible yet because not enough data is available to perform a meta-regression, or the link is weaker than with classical chemotherapies, tumor response based rules may not be optimal.
In this paper, we explore the use of a multistate model for decision making based on single-arm early phase trials. The multistate model allows to account for more information than the simple RECIST response status, namely, the time to get to response, the duration of response, the PFS time and time to death. We propose to base the decision on efficacy on the OS hazard ratio (HR), predicted from a multistate model based on early phase data with limited survival follow-up, combined with historical control data. Using three case studies and simulations, we illustrate the feasibility of the estimation of the OS HR using a multistate model based on limited data from early phase studies. We argue that, in the presence of limited follow up and small sample size, and on assumptions within the multistate model, the OS prediction is acceptable and may lead to better decisions for continuing the development of a drug.}
\Plainkeywords{Multistate Model, Clinical Trial, Decision Making} 

\Address{
  F. Hoffmann-La Roche AG\\
  Grenzacherstrasse 124
  4070 Basel, Switzerland\\
  E-mail: \email{ulrich.beyer@roche.com}\\
  URL: \url{http://www.roche.com}
}

\usepackage{amsmath}
\usepackage{amssymb}
\usepackage{amsfonts}
\usepackage{amsthm}
\usepackage{latexsym}
\usepackage{thumbpdf}
\usepackage{color}

\newtheorem{theorem}{Theorem}[section]

\input{commands.tex}

\begin{document}


\section{Introduction}
\label{intro}

\noindent In pharmaceutical oncology drug development, single-arm trials are often used in early phase to gather the first evidence of a new molecule's efficacy, with drug activity determined through tumour response using the RECIST criterion \citep{brown_11, grayling_16}. The observed complete response proportion (CR), overall response proportion (OR), defined as CR + partial responders, or disease control proportion (DC) defined as OR + stable disease patients, is compared to a benchmark  based on historical control data. A decision is then made whether to advance the molecule into pivotal randomized clinical trial testing, or Phase 3. The implicit assumption for this gating approach is, that the measures evaluated at the end of the early development phase are good surrogates for the primary endpoint in Phase 3, either progression-free (PFS) or overall survival (OS). Such surrogacy is generally assessed through meta-analytic methods \citep{buyse_00}, but in many indications considered to be moderate to poor \citep{fiteni_16, nakashima_16, kemp_17}.

With the advent of immunotherapy in oncology, surrogacy associations between response and OS are further complicated by three aspects: First, in many indications there is only very limited data available yet, and/or OS data is not mature enough, to explore the association between effect measures on a trial level. Second, the immunotherapy mechanism of action (MoA) insinuates that the association between a response-based endpoint and OS might be different from previous chemo- or targeted-therapy based agents. \cite{mushti_18} provide an early analysis of 13 studies on a new class of agent in five indications. Their conclusion is, that responders exhibited longer survival and PFS than non-responders, but that the trial- and individual level associations were weak between ORR or PFS and OS. Note that the statement about the longer OS for responders might potentially be prone to immortal bias \citep{weber_18}. Third, immunotherapy leads to a delayed effect of the therapy on PFS or OS \citep{ref:Hoos2012, ref:Chen2013}. These papers highlight the challenges of the classical response and OS analysis linked to the delayed effect of the immunotherapies. The consequences of the delayed effect are that response patterns are different and that OS shows a non-proportional ratio when compared to classical chemotherapies.

The immunotherapy MoA suggests that compared to standard chemo- or targeted therapy, immunotherapy potentially acts by not only increasing the proportion of responders, but also prolonging response duration (duration of response, DOR) for responding patients and post-progression survival, potentially implying increased OS. Taken together with the weak surrogacy association in general, this raises the question of whether gating criteria in early decision making in drug development can be improved by considering more granular information. Ideally, one would like to decide on whether to advance a molecule in Phase 3 not only based on comparing the estimated response proportion in a cohort treated with the new drug to an estimated proportion received from a historical control cohort, but get an unbiased {\it estimate} of the OS survival function for the new molecule.

Now, early phase trials generally are neither large enough nor do they have long follow-up. As a consequence, little information on OS can be collected directly from the trial. Instead of estimation from data, one thus needs to {\it predict} the OS survival function. In this paper, we propose to compute such a prediction via a multistate model that borrows estimates of transition hazards from another trial in the same population. This prediction of the survival function in the experimental arm can then be benchmarked to n estimate of the survival function based on the control. The latter is generally available from previous trials and/or historical data. Based on this comparison, operating characteristics of a potential Phase 3 trial of the new molecule can then be explored, and these can be compared to their counterparts based on response only.



In this multistate model, transition hazards and probabilities  can either be estimated based on the already available  data from the early phase trial of the new molecule itself, or (some of them) can be borrowed from the historical control trial, under plausible assumptions.

Since in a first step, our aim is to explore the general feasibility of this approach, we will put the concept to test by predicting OS based on early phase information {\it retrospectively} in three randomized Phase 3 clinical trials (RCT) in three different disease indications (breast, bladder, and non-small cell lung cancer). In these RCTs, artificially administratively censored data in the experimental arm reflects early phase data from the new molecule, and the entire control arm data corresponds to the historical control. After fitting a multistate model, we explore operating characteristics as a function of an assumed gating value of the OS hazard ratio. Furthermore, we provide a recommendation on the minimal survival follow-up necessary to obtain a robust estimate of the multistate model to support robust decision-making.

Usually, the clinical relevance of a difference in response proportions is only indirectly accessible at best, e.g. through meta-analyses. Having a reliable OS prediction for the new molecule also enables to compute an effect measure, the OS HR, that is relevant for how the {\it patient feels, functions, and survives} (\citealp{temple_95}).

The proposed setup of taking the {\it historical control data} from the control arm of an RCT provides an idealized situation to test our multistate model on, since it avoids issues with both, selection and calendar time bias. For a new general gating approach to be valid at all it has to work in this ideal scenario. A discussion of how to potentially address the bias questions is provided in Section~\ref{summary}.

This paper is structured as follows: In Section~\ref{casestudies}, we describe our motivating examples that highlight the challenges posed by classical decision making. In Section~\ref{methods}, we introduce the proposed multistate model whereas in Section~\ref{sec:haz_same}, we illustrate the feasibility of the proposed approach under the assumption that post-progression information can be borrowed from historical data. In Section~\ref{sec:haz_diff}, we illustrate the feasibility of the proposed approach using a less restrictive assumption. We end this paper by a summary and conclusion in Section~\ref{summary}.

\section{Case studies}
\label{casestudies}

In this section we briefly describe the three RCTs from which we will use the data to develop the multistate approach.

\subsection{Cleopatra trial: New antibody as add on to chemotherapy}

Cleopatra was a Phase 3 RCT which compared Docetaxel + Trastuzumab + Pertuzumab (experimental treatment) against Docetaxel + Trastuzumab + Placebo (control treatment) in 808 previously untreated HER2-positive metastatic breast cancer patients using a primary endpoint of independently-assessed PFS \citep{baselga_12}. Here, Docetaxel is a chemotherapy and Trastuzumab a targeted-therapy agent with an antibody dependent cellular cytotoxicity (ADCC) mode of action. As soon as progression was diagnosed for a patient, treatment was stopped, in both arms. At the primary cutoff, in the intention-to-treat (ITT) population, the combination arm prolonged PFS (hazard ratio, HR = 0.62, 95\% confidence interval, CI, from 0.51 to 0.75), OS (HR = 0.64, 95\% CI from 0.47 to 0.88), and increased the proportion of responders by 0.108 from 0.693 to 0.802. In a follow-up analysis \citep{swain2015pertuzumab}, prolongation of median DOR by 7.7 months was also reported.

\subsection{Imvigor211 trial: Immuno- versus chemotherapy}

Another Phase 3 RCT, Imvigor211, compared Atezolizumab (experimental treatment) against chemotherapy (control treatment) in patients with platinum-treated locally advanced or metastatic urothelial carcinoma \citep{powles2017atezolizumab}. Atezolizumab is a cancer immunotherapy. In this trial, treating patients post-progression in the Atezolizumab arm was allowed, but not mandatory. The decision whether to continue treatment was at the discretion of the investigator. In the ITT population, this trial showed an overall benefit in OS (HR = 0.84, 95\% CI from 0.72 to 0.97). Note that the OS survival functions crossed and Atezolizumab benefit was only seen after approximately 7 months. Neither response proportion nor PFS showed a benefit in favor of Atezolizumab, even more progressors were reported in the Atezolizumab group. However, in the responding patients, DOR was prolonged from a median of 7.4 months in the chemotherapy to 21.7 months in the Atezolizumab arm.

\subsection{Oak trial: Immuno- versus chemotherapy}

Finally, Oak was another Phase 3 RCT comparing Atezolizumab (experimental treatment) versus Docetaxel (control treatment) in patients with previously treated non-small-cell lung cancer \citep{rittmeyer2017atezolizumab}. Similar to Imvigor211, post-progression treatment was allowed in the Ateolizumab arm. Results were also comparable to Imvigor211: In the ITT population, an OS HR of 0.73 (95\% CI from 0.62 to 0.87) was observed, with no benefit for either response or PFS. As in Imvigor211, DOR was prolonged in the Atezolizumab arm with a median of 16 months versus 6.2 months in the Docetaxel arm.

Tables~\ref{tab:cleopatra}-\ref{tab:oak} in the appendix provide a detailed summary of the results of the three case studies.

The Imvigor211 and Oak trials illustrate the difficulty of making early drug development decisions based on response proportions when ultimately in Phase 3, a long-term endpoint such as OS is assessed. In both trials, response proportions comparing arms were virtually identical, but the OS HR point estimate and upper limit of a 95\% confidence interval was below 1.

\section{Methods}
\label{methods}

\subsection{Multistate models in general}

Classical survival analysis focuses on the time from some time origin until the occurrence of an event
of interest, a quantity typically denoted {\it survival} or {\it failure time} (\citealp{beyersmann_12}).
Further events prior to the event of interest may substantially change the risk of the event of interest to
occur, e.g. response or progression may change the risk of death in an oncology setting. If collected, such intermediate events provide more detailed information on the disease process and allow for more precision in predicting the event-time of a patient. Multistate models provide a framework that allow for the analysis of such event history data. Figure \ref{fig:MM} provides an example of a multistate model that connects the three canonical oncology endpoints, namely response, progression, and death. In total, in this model six transitions are possible.

For a detailed introduction into multistate models we refer to standard references such as \cite{putter2007tutorial}, \cite{andersen_08}, \cite{aalen_08}, or \cite{beyersmann_12}. Here, we only give a very brief description to allow definition of the quantities relevant for this paper. Let $(X_t)_{t \ge 0}$ denote a stochastic process with state space $\{1, \ldots, J\}$, where $X_t$ denotes the state an individual is in at time $t$. {\it Events} are then modelled as transitions between the states of the model, so a progression out of the initial SD state in Figure~\ref{fig:MM} would be modelled as a $1 \to 3$ transition. A key quantity in multistate modelling are transition probabilities, defined as
\bea
  P_{ij}(s, t) &:=& P(X_t = j |X_s = i, \text{Past}), s \le t.
\eea
The quantity $P_{ij}(s, t)$ denotes the probability to be in state $i$ at time $s$ and in state $j$ at time $t$, where {\it Past} is a $\sigma$-algebra generated by the process $X$ up to time $s$, or more informally the accumulated knowledge about the process' history up to $s$. A transition hazard, i.e. the instantaneous risk to move from state $i$ to $j$ at time $t$, is defined as
\bean
\lambda_{ij}(t) \cdot d t &:=& P(X_{(t+dt)}- = j |X_{t-} = i), i,j \in {1, 2, . . . , J}, i \ne j,
\label{eq: transition hazard}
\eean
and the corresponding cumulative transition hazard by
\bea
\Lambda_{ij}(t) &=& \int_0^t \lambda_{ij}(s) ds. \label{eq: cumulative transititon hazard}
\eea
In what follows we will assume $X$ to be Markov.
This assumption appears sensible in our specific oncology application, since a patient with an early tumor progression will likely have a different subsequent risk to die compared to a patient being progression-free for a longer period.

\subsection{OS prediction in a four-state oncology multistate model}
\begin{figure}[htb]
\begin{center}
\caption{Oncology multistate model used for OS predictions.}
\includegraphics{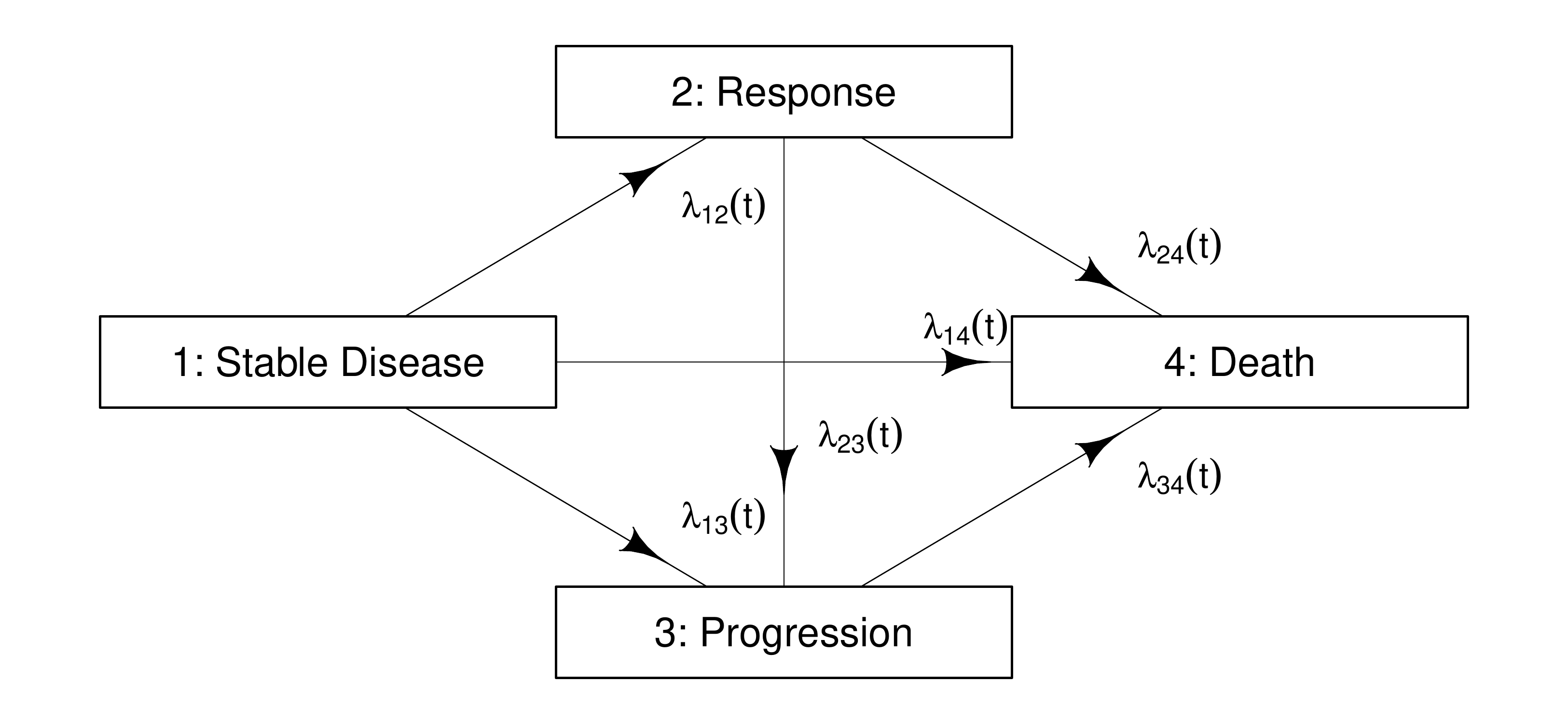}
\label{fig:MM}
\end{center}
\end{figure}

In Sections~\ref{sec:haz_same} and \ref{sec:haz_diff} we will use a multistate model to make OS predictions. The model is depicted in Figure~\ref{fig:MM} and consists of four states: patients enter having stable disease (State 1, SD). Out of this initial state, after receiving treatment, they can either respond (Response, State 2), progress (PD, State 3), or die without observed response or progression (Death, State 4). Patients responding can either progress or die, and patients who progressed remain at the risk of dying, i.e. a $3 \to 4$ transition.

In order to inform decision-making in early oncology drug development as described in Section~\ref{intro} not only based on response, but on OS predictions based on the above multistate model, we have to derive the corresponding survival function for OS, $S_{OS}$. This function will depend on the transition hazards $\lambda_{ij}$ and/or the cumulative versions $\Lambda_{ij}$, either directly or indirectly through transition probabilities $P_{ij}$, for $i, j \in \{1, \ldots, 4\}$. Specifically,
\bean
S_{OS}(t) &=& 1-P_{14}(0,t) \nonumber \\
&=& 1 - \Bl(P_{1  \to 4}(0,t) + P_{1  \to 3  \to 4}(0,t) + P_{1  \to 2  \to 4}(0,t) +  P_{1  \to 2  \to 3  \to 4}(0,t) \Br). \label{eq:S_overall}
\eean
The four terms $P_{1  \to 4}, P_{1  \to 3  \to 4}, P_{1  \to 2  \to 4},$ and $P_{1  \to 2  \to 3  \to 4}$ quantify the transition probabilities to move from the initial State 1 to the death State 4, via all possible paths. The corresponding formulas are derived in the appendix, and how to estimate $S_{OS}$ is discussed in Section~\ref{subsec:model_ass}.

\subsection{Model assumptions}
\label{subsec:model_ass}

We will use a multistate model to compute a prediction of $S_{OS}$ for the patients recruited to an {\it early phase} trial. In what follows, the characteristics of such an early phase trial are:
\bi
\item[i] Single-arm trial with a modest number of patients, $n$. We assume throughout that $n = 40$.
\item[ii] Limited follow-up for post-progression survival, as described below, making estimation of the survival function for OS impractical.
\ei
In oncology, early development decisions are often based on this type of trials, potentially with a different number of patients. In such trials, because an OS assessment is typically infeasible, the focus traditionally lies on an assessment of response proportions. For immunotherapeutic treatments, DOR also gained in popularity as an effect measure. Thus, information on the transitions out of State 1 into States 2 and 3 (whether through State 2 or direct), and the corresponding censoring information for patients remaining in State 1 or 2, is systematically collected. This enables unbiased estimation of the transition hazards $\lambda_{12}, \lambda_{13}$, and $\lambda_{23}$. The transitions $1 \to 4$ and $2 \to 4$ rarely happen at all, and typically no relevant difference is seen between arms. If they happen, the corresponding death dates are typically collected within an early phase trial, because patients are generally followed-up until progression. As a result, we get little, but completely collected event data to estimate these transition hazards. Since these transitions are rare, the resulting transition probabilities are generally low, implying that their contribution to the value of $S_{OS}$ is very limited.

For the transition hazard $\lambda_{34}$, or {\it post-progression hazard}, data collection in early phase trials is much less systematic, because protocol-specified follow-up generally ends after progression. This implies that once a patient progresses, the information about a potential later death, i.e. the $3 \to 4$ transition, and corresponding last-known alive dates are only anecdotically known, at best. This makes estimation of $\lambda_{34}$ challenging, if not impossible, at the time when a decision needs to be taken whether to take a molecule forward into Phase 3. This, because - as opposed to the transitions $1 \to 4$ and $2 \to 4$ where we collect little but (virtually) complete data - we only observe a small proportion of the death events that actually happen after progression.

So, instead of {\it estimating} the OS survival function of the experimental treatment we will {\it predict} it based on the multistate model. To be able to do this, we need three ingredients:
\bi
\item[i] {\it Historical data} to inform the prediction, i.e. data sampled from a population that does not systematically differ from the one the early phase trial is drawn from. Our case studies below provide an idealized scenario for this: since all of them are pivotal RCTs, we can simply use the control arm as our historical data. However, in a real-world application of the proposed methodology, the historical control data will typically be external to the early phase trial, implying a potential risk of selection or calendar time bias. We will discuss this point further in Section~\ref{summary}.
\item[ii] The historical data needs to have (much) longer and more comprehensive survival follow-up compared to the early phase trial. Again, the idealized scenario of using RCTs as proof-of-concept provides this.
\item[iii] We need to make an assumption about the post-progression hazard functions in both treatment arms. In Section~\ref{sec:haz_same} the one in the experimental arm will be completely borrowed from the historical control whereas in Section~\ref{sec:haz_diff}, we will generalize this assumption to the two post-progression hazards not being identical, but proportional.
\ei

%
To compute $S_{OS}$, in our case studies reported in Sections~\ref{sec:haz_same}-\ref{sec:minFU}, we use cumulative hazard functions predicted from Cox regressions, with a binary covariate for treatment arm. In contrast to treatment-specific Nelson-Aalen estimates of the cumulative hazard for each transition, this has the following advantages: it allows to receive predictions up to the last available observation (censored or event) in {\it both} treatment arms. As we will compare a historical control arm with generally long-term follow-up to an experimental arm with much less follow-up, having a predicted survival function up to the last observation in the control arm is helpful. Furthermore, using Cox regression we either simply use an estimated baseline hazard from the control arm (when assuming identical post-progression hazards in Section~\ref{sec:haz_same}) or an estimate based on the data of both arms (proportional post-progression hazards in Section~\ref{sec:haz_diff}), i.e. in both cases we borrow strength from the control arm to make a prediction in the experimental arm. Finally, using Cox regression to predict cumulative hazard functions also easily allows for inclusion of (transition-specific) covariates. For the different transitions, we will keep the model maximally flexible by assuming that each transition has its own baseline  hazard.

So, within our multistate model framework, the hazard for a $i \rightarrow j$ transition for a subject with a covariate vector $Z$ is given by
\bean
\lambda_{ij}(t \,|\,Z) &=& \lambda_{ij,0}(t) \cdot \exp(\beta_{ij}^T Z). \label{eq: Cox PH}
\eean
Here, $\lambda_{ij,0}(t) $ is the baseline hazard for the considered transition and $\beta_{ij}$ the vector of regression coefficients. In the case studies in Sections~\ref{sec:haz_same}-\ref{sec:minFU} treatment information, i.e. whether a patient belongs to the historical control or the experimental arm, will be the only covariate in our transition-specific Cox models. As a consequence, $\beta_{ij}$ describes the treatment effect on the $i \rightarrow j$ transition hazard. From a hazard function estimate $\hat \lambda_{ij}(t \,|\,Z)$ we can then easily predict the cumulative hazard function in both treatment arms, and these predictions are plugged-in into \eqref{eq:S_overall} to get the overall survival function estimate.

Although we do not pursue this here when discussing early decision-making, further covariates, even transition-specific, can be incorporated in the prediction of the cumulative hazard functions.

Variability of $S_{OS}$ can be assessed as described in \cite{ref:Wreede2010}.

\section{OS prediction if post-progression hazards are assumed identical}
\label{sec:haz_same}

Under the assumption of identical post-progression hazards in the control and the experimental arm, the transition hazard $\lambda_{34}(t \,|\,Z)$ does not depend on the treatment covariate, i.e. $\lambda_{34}(t \,|\,Z) = \lambda_{34}(t) = \lambda_{34,0}(t)$. Under this assumption, a prediction based on a multistate model is simplified, and the post-progression information in the experimental arm can be borrowed completely from the control arm.

The assumption of identical post-progression hazard is applicable  when the treatments under investigation are not expected to provide benefit beyond progression. Indeed, the chemotherapy as well as ADCC mode of action work via a direct tumor cell killing mechanism. In both cases, treatment will be stopped after a diagnosis of progression, the treatment does no longer work and it seems reasonable that patients progressing at a same time will have similar transition hazards from progression to death, regardless of whether they received experimental or control therapy.

In Cleopatra, both arms belonged to the ADCC treatment class. In the following section, we justify that the assumption of similar post-progression hazards is valid and illustrate that the multistate model can predict long term survival of the experimental arm with limited post-progression death events in the experimental arm. We first show that the multistate model, based on the Cox model and with the assumption of equal post-progression hazards, can reproduce the raw Kaplan-Meier (KM) estimates based on the full dataset. Second, we show that the multistate model can provide a reliable prediction of the OS survival function in the experimental arm when those patients who experience a post-progression death are censored one day after their progression date (hence, the hazard estimation is solely based on post-progression data from the control arm). Third, we explore operational characteristics of an early development decision based on the predicted OS in a typical early phase setting via simulating from the original data.

\subsection{Analysis of the full dataset}


Estimating the transition hazards in the multistate model described in Figure~\ref{fig:MM} using Cox regression with treatment as a covariate yields the estimated hazard ratios in Table~\ref{cleofull}. The number of events for each transition are reported in Table~\ref{tab:Cleotrans} in the appendix. The estimated hazard ratio of the transition from progression into death amounts to 0.97, so that the assumption of equal post-progression hazard functions therefore appears justified.

\begin{center}
\begin{table}[h!]
\centering
\caption{Estimated transition hazard ratios from Cleopatra based on full dataset.}
\label{cleofull}
\begingroup\footnotesize
\begin{tabular}{cccr}
  \hline
transition & hazard ratio & 95\% CI & $p$-value \\
  \hline
$1 \to 2$ & 1.16 & [0.98, 1.37] & 0.09 \\
  $1 \to 3$ & 0.64 & [0.47, 0.88] & 0.006 \\
  $1 \to 4$ & 0.66 & [0.28, 1.55] & 0.34 \\
  $2 \to 3$ & 0.70 & [0.55, 0.89] & 0.0041 \\
  $2 \to 4$ & 0.21 & [0.06, 0.78] & 0.02 \\
  $3 \to 4$ & 0.97 & [0.68, 1.37] & 0.85 \\
   \hline
\end{tabular}
\endgroup
\end{table}\end{center}

\medskip

We can now predict the OS survival function in the experimental arm under the assumption of common transition hazards from progression to death. Figure~\ref{CleopatraFullData} displays the KM estimate based on the original full dataset, the prediction based on the multistate model, also based on the full dataset. We find that the multistate model-based prediction is accurate under the assumption of proportional transition hazards and common transition hazard from progression to death, i.e. the $3 \to 4$ transition.

\subsection{Censoring of deaths immediately after progression in the experimental arm}
\label{subsec: censpd}

\begin{figure}[!ht]
\begin{center}
\caption{Cleopatra: KM estimates and multistate (MS) predictions of $S_{OS}$.}
\setkeys{Gin}{width=1\textwidth}
\includegraphics{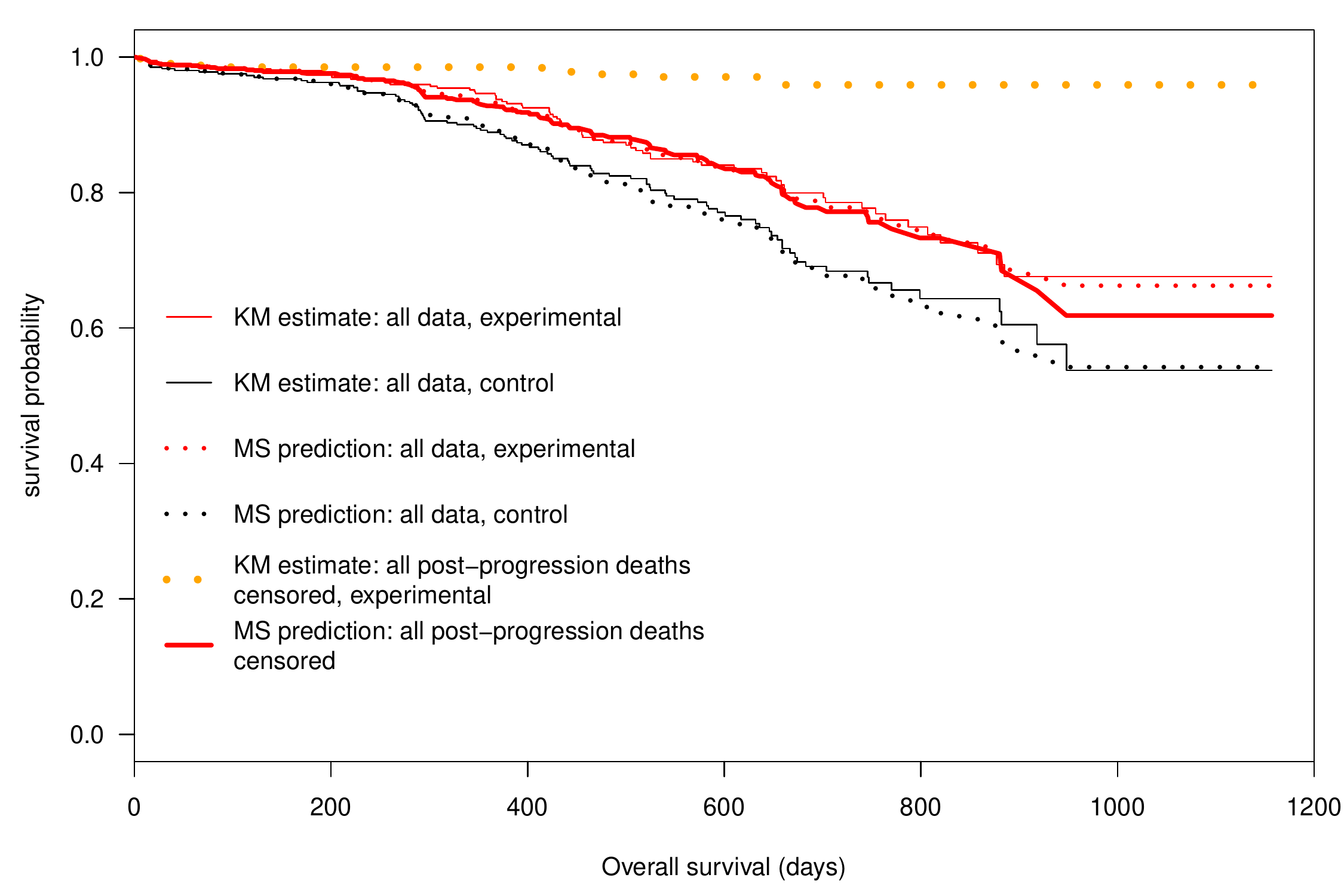}
\label{CleopatraFullData}
\end{center}
\end{figure}

As a next step, we illustrate that the approach based on the multistate model is capable of estimating the OS survival function, even if all post-PD events in the experimental arm are censored one day after progression. In this case, the estimate of the transition hazard from progression to death is based on data from the control arm only. The corresponding KM estimate of the experimental arm is provided in Figure~\ref{CleopatraFullData} (orange line). Including only deaths without progression illustrates the very limited OS information that enters the estimation of the multistate model, see Table~\ref{tab:Cleotrans}. Still, multistate models are able to accurately predict the OS survival function, even in absence of post-progression data, the typical situation in early phase clinical trials as described in Section~\ref{subsec:model_ass}. This is illustrated by the bold red line in Figure~\ref{CleopatraFullData}.

\subsection{Simulation of operating characteristics for an early phase trial}

Finally, we want to assess whether the OS prediction of the experimental arm, based on limited early data from this experimental arm and the post-progression hazard completely from the control arm, indeed yields a decision rule with good operational characteristics. To this end, the experimental arm in the Cleopatra trial represents data from the new molecule to be tested in the early phase trial, and the control arm reflects the (historical) comparator data. We mimic a typical early phase setting using simulation as follows:

\begin{enumerate}
\item[1] In the control arm, all patients are used for estimation.
\item[2a] To mimic the experimental arm assuming the new drug does not work, an early phase trial with 40 patients will be sampled with replacement from the control arm again. This allows assessment of the false-positive rate.
\item[2b] Assuming the experimental drug works as in Cleopatra, forty patients are sampled with replacement from the Cleopatra experimental arm, in order to estimate the false-negative rate.

In both, 2a and 2b, post-progression deaths for the sampled patients are censored immediately after progression. $S_{OS}$ is then predicted for the 40 patients sampled from the experimental arm, based on the multistate model and the full control arm.

\item[3] The hazard ratio comparing experimental to control is approximated by fitting an exponential survival function to the predicted OS survival functions and taking the ratio of estimated exponential hazard parameters.
\item[4] Based on these hazard ratio estimates, we derive operating characteristics of a decision defined as follows: a decision to further develop the experimental arm is made when the observed hazard ratio is below a target value. These target values are chosen to be 0.8, 0.85, 0.9 and 1. Operating characteristics are estimated based on 1000 simulated early phase trials of 40 patients by computing the proportion of simulations that yield a hazard ratio above (false-negative) or below (false-positive) the considered target value.
\end{enumerate}

In the simulations, recruitment was generated such that the 40 sampled patients were uniformly entering the trial during 6 months. The analysis is performed 9 months after the last enrolled patient, such that the longest possible observation time is up to 15 months for the first patient, and less for those recruited later. The follow up of 9 months after last patient randomized was chosen based on the following consideration: median PFS time for the control arm was about 12 months. To allow for enough time to collect a sufficient number of progression events in the experimental arm, even with a prolonged time to progression, follow-up time was set to 15 months.

\begin{figure}[!ht]
\begin{center}
\caption{Results of one of 1000 simulated early phase trials of 40 patients, sampled from the experimental arm.}
\setkeys{Gin}{width=1\textwidth}
\includegraphics{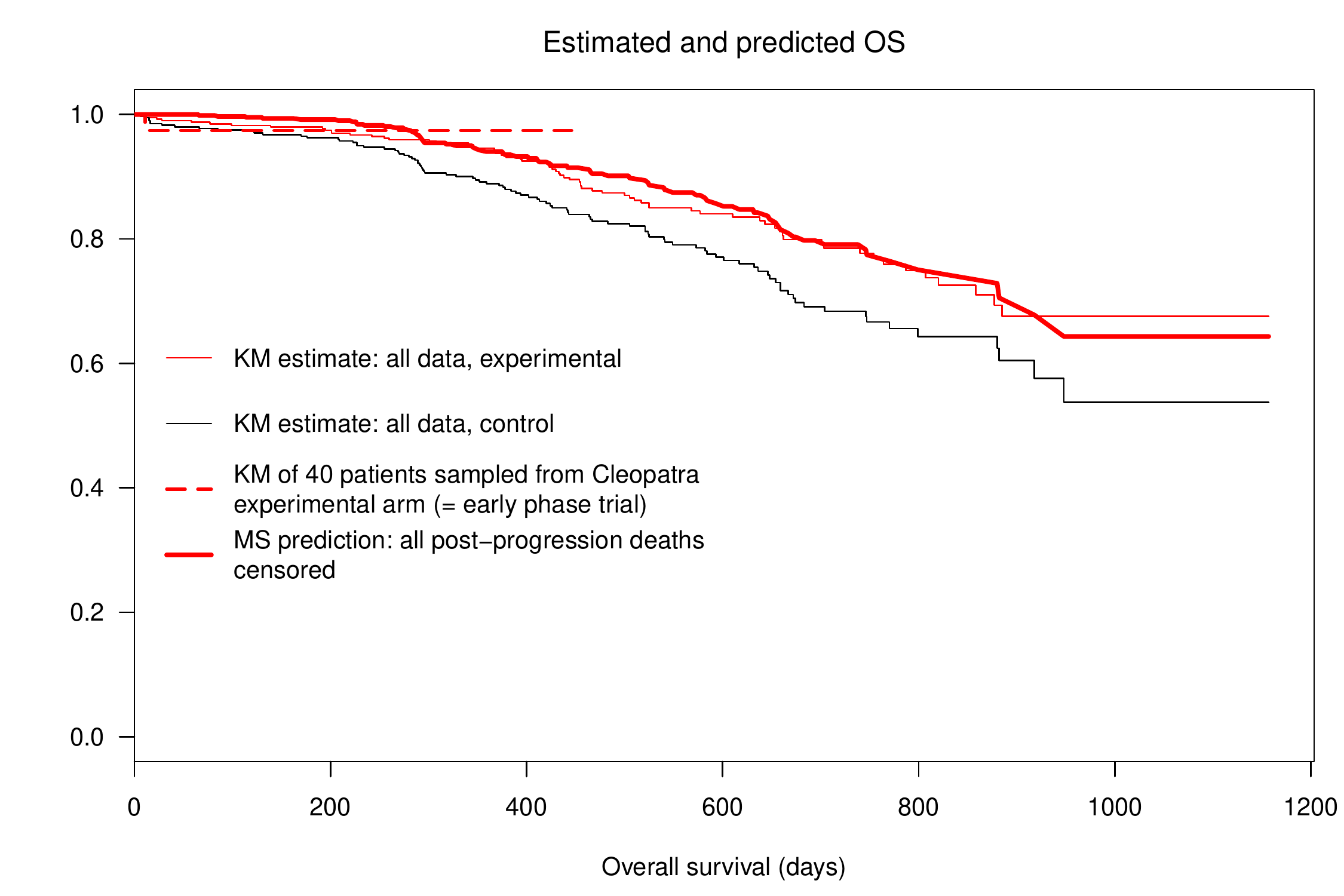}
\label{CleopatraSIM1}
\end{center}
\end{figure}

For illustration, the results of one simulation run when sampled from the experimental arm are provided in Figure~\ref{CleopatraSIM1}, with the first plot showing the KM estimate of the OS survival function of the 40 early phase trial patients (solid) and the KM estimate of the experimental arm survival function of Cleopatra (dashed). As expected, the early phase trial OS data is immature and not able to provide a good estimate of the OS survival function. The second plot again shows the estimated survival function from Cleopatra, now for the control and experimental arm (black and red, respectively) in addition to the predicted OS survival function for the experimental arm from the multistate model (solid red line), based on the 40 patients. We find that the multistate model prediction appears to be able to accurately capture the shape of the Cleopatra experimental arm survival function. This illustrates how the multistate model borrows information from the control arm for the estimation of the OS survival function on the experimental arm.

\begin{figure}[!ht]
\begin{center}
\caption{Multistate estimation of the OS survival function based on 1000 simulated early phase trials of 40 patients (red lines) sampled from the experimental arm (left) and from the control arm (right) and using the complete control arm from the Cleopatra trial. Black lines represent the KM estimate of OS based on the full experimental (solid) and control (dashed) arm.}
\setkeys{Gin}{width=1\textwidth}
\includegraphics{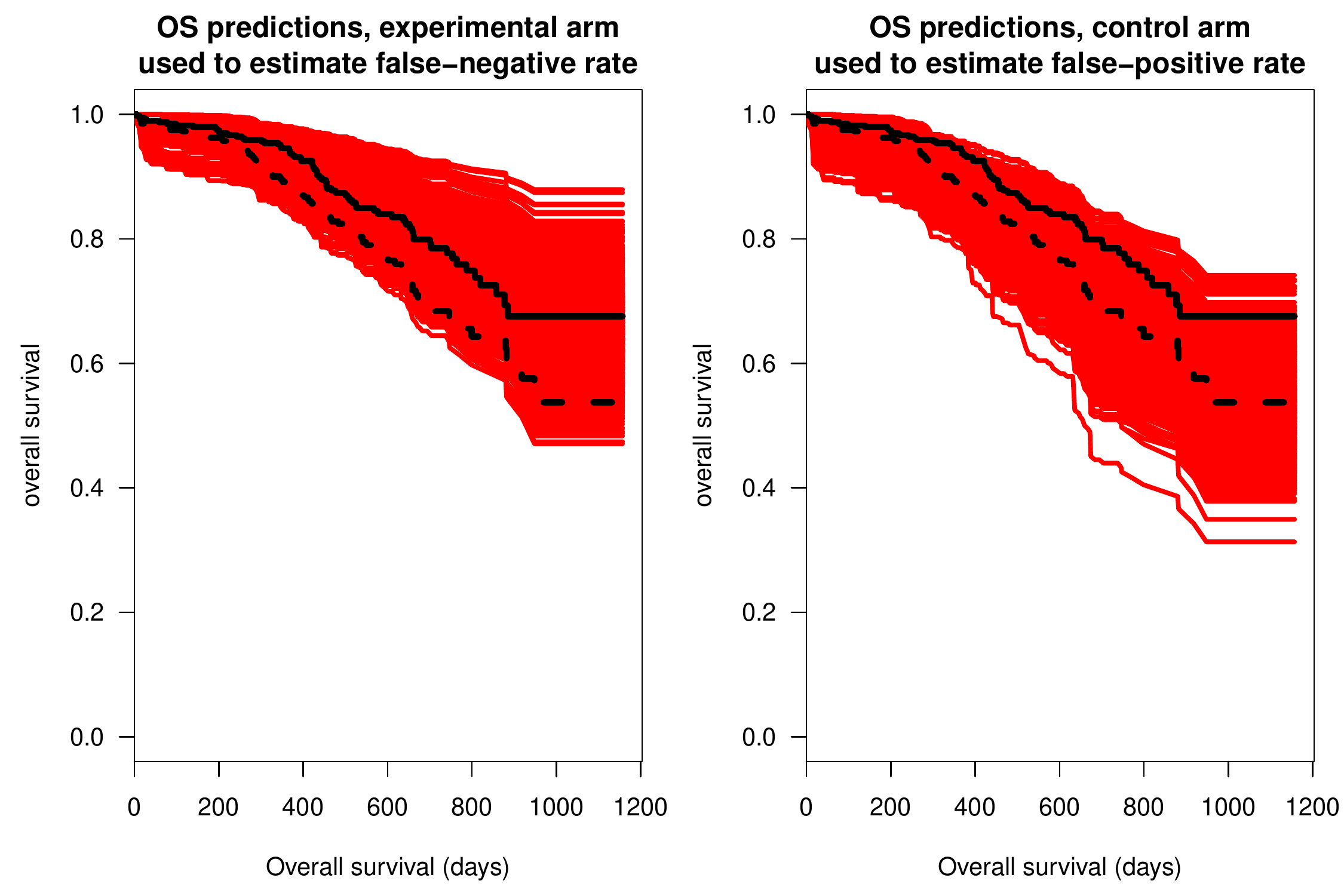}
\label{CleopatraSIMall}
\end{center}
\end{figure}

Results of multistate OS estimation from the 1000 simulated early phase trials sampled from the experimental (left) or control (right) Cleopatra arm are given in Figure~\ref{CleopatraSIMall}. The false-negative and false-positive rate of the decision rule defined above are given in Table \ref{CleopatraSIMtable}. The operating characteristics are favourable in an early phase trial setting in oncology, and illustrate the feasibility of a decision rule using a small trial with limited data (restricted follow up and low number of post-progression deaths) based on a multistate model, where the transition from progression to death is borrowed from the historical control data. Furthermore, operating characteristics are based on a prediction of OS, instead of response, whose quality as a surrogate and relevance for the patient is often questionable, as discussed in Section~\ref{intro}.

The mean HR over the 1000 simulated early phase trials amounts to 0.62 for the left table, matching well the observed OS HR in Cleopatra (see Table~\ref{tab:cleopatra}). For the right table we get 0.99, again accurately reflecting the assumed scenario of no effect.

\begin{figure}
\caption{table}{False-negative and false-positive rate of multistate decision rule based on 1000 simulated early phase trials of 40 patients sampled from the experimental (left) or control (right) arm in Cleopatra.}
\label{CleopatraSIMtable}
\begin{center}
\begin{tabular}{lcc}
\multicolumn{3}{c}{False-negative rate} \\
& HR target value & \% HR $>$ target value  \\
\hline & $0.80$&0.16 \\
 & $0.85$&0.11 \\
 & $0.90$&0.07 \\
 & $1.00$&0.04 \\
 & $0.80$&0.16 \\ \hline\end{tabular}
\end{center}
\begin{center}
\begin{tabular}{lcc}
\multicolumn{3}{c}{False-positive rate} \\
 & HR target value &  \% HR $<$ target value  \\
\hline & $0.80$&0.21 \\
 & $0.85$&0.29 \\
 & $0.90$&0.39 \\
 & $1.00$&0.55 \\
 & $0.80$&0.21 \\ \hline\end{tabular}
\end{center}
\end{figure}

\section{OS prediction if post-progression hazards are assumed proportional}
\label{sec:haz_diff}

Under the assumption of proportional post-progression hazard, the transition $\lambda_{34}(t \,|\,Z)$ depends  on the treatment covariate, i.e. $\lambda_{34}(t \,|\,Z) =   \lambda_{34,0}(t) \exp(\beta_{34} Z)$. The transition hazard in the experimental arm is assumed to be proportional to the one in the control arm. Under this assumption, the baseline hazard $\lambda_{34,0}(t)$ is estimated from both arms, and there should be sufficient data on the experimental arm to allow estimation of $\beta_{34}$. We address the question of what {\it sufficient data} could mean in Section~\ref{sec:minFU}.

This assumption is plausible if the mode of action of the experimental arm is expected to provide benefit beyond progression.The goal of cancer immunotherapy (CIT) is to boost or restore the ability of the immune system to detect and destroy cancer cells by overcoming the mechanisms by which tumors evade and suppress the immune response (\citealp{disis_14}). This implies that the patient may still benefit of the treatment even after the disease progresses. For this reason, in trials assessing CIT, often treatment beyond progression is allowed by the protocol in the CIT arm, as long as the investigator considers the treatment still to provide clinical benefit.

Both the Imvigor211 and Oak trial compared a CIT to a chemotherapy. Given the mode of action of the experimental treatment and the fact that treatment beyond progression was allowed, the transition hazards from progression to death must be assumed to be different. Rather than leaving the transition hazards from progression to death completely unspecified, we assumed proportionality of these hazards between treatment arms. The benefit of this approach is to borrow from the control some information on the shape of the transition hazards.
We show later that this assumption is plausible on both trials.

In this section, we replicate the investigations from Section~\ref{sec:haz_same} on the Imvigor211 and Oak data. First, we assess the plausibility of the proportional post-progression hazards assumption on the full dataset for both studies. Second, we show that the full dataset with limited follow up allows to recover the HR for OS as observed on the full dataset. Third, we simulate early phase trials with limited follow up and illustrate that the multistate model can be used for decision making.

\subsection{Analysis of the full dataset for Imvigor211 and Oak}


Tables~\ref{imvigorfull} and \ref{oakfull} provide estimated hazard ratios for each transition for the two studies. Number of events for each transition are reported in Tables~~\ref{tab:Imvigortrans} and \ref{tab:Oaktrans} in the appendix. We find evidence that the assumption of post-progression transition hazard functions being equal appears unrealistic, unlike in Cleopatra discussed in Section \ref{sec:haz_same}. A statistical test that assesses the null hypothesis of proportionality of hazards (\citealp{grambsch_94}) gives $p$-values of 0.78 and 0.93, respectively. The proportional hazards (PH) assumption for the post-progression hazards thus appears reasonable in both studies.

\begin{center}
\begin{table}[h!]
\centering
\caption{Estimated transition hazard ratios from Imvigor211 based on full dataset.}
\label{imvigorfull}
\begingroup\footnotesize
\begin{tabular}{cccr}
  \hline
transition & hazard ratio & 95\% CI & $p$-value \\
  \hline
$1 \to 2$ & 1.22 & [0.85, 1.73] & 0.28 \\
  $1 \to 3$ & 1.40 & [1.19, 1.65] & < 0.0001 \\
  $1 \to 4$ & 0.95 & [0.70, 1.29] & 0.74 \\
  $2 \to 3$ & 0.36 & [0.21, 0.62] & 0.00024 \\
  $2 \to 4$ & 0.16 & [0.03, 0.75] & 0.02 \\
  $3 \to 4$ & 0.72 & [0.60, 0.85] & 0.00019 \\
   \hline
\end{tabular}
\endgroup
\end{table}\end{center}

\begin{center}
\begin{table}[h!]
\centering
\caption{Estimated transition hazard ratios from Oak based on full dataset.}
\label{oakfull}
\begingroup\footnotesize
\begin{tabular}{cccr}
  \hline
transition & hazard ratio & 95\% CI & $p$-value \\
  \hline
$1 \to 2$ & 1.10 & [0.76, 1.59] & 0.61 \\
  $1 \to 3$ & 1.28 & [1.08, 1.51] & 0.0046 \\
  $1 \to 4$ & 0.58 & [0.40, 0.84] & 0.0041 \\
  $2 \to 3$ & 0.39 & [0.24, 0.65] & 0.00028 \\
  $2 \to 4$ & 0.16 & [0.03, 0.79] & 0.02 \\
  $3 \to 4$ & 0.80 & [0.66, 0.96] & 0.02 \\
   \hline
\end{tabular}
\endgroup
\end{table}\end{center}

\subsection{Analysis on data with limited follow up}

In this section, we  confirm that the multistate model is able to provide a valid estimate for the OS HR on data with limited follow up. In both trials, we censored the post-progression death events that occurred later than 180 days after randomization in the experimental arm of both studies. This corresponds to typical minimal follow up of early phase studies.

\begin{figure}[!ht]
\begin{center}
\caption{Estimation of the OS survival function in Imvigor211 (left) and Oak (right) when death events are censored after 180 days after randomization (indicated through the vertical line). The orange line is the survival function estimate of the experimental arm, where all post-PD events after 180 days are administratively censored. The black lines depict the KM estimates of OS on the full dataset for the control arm. The green line is the survival prediction based on an estimated HR from a Cox regression model computed on both arms, and the red line is the survival prediction based on the multistate model.}

\setkeys{Gin}{width=1\textwidth}
\includegraphics{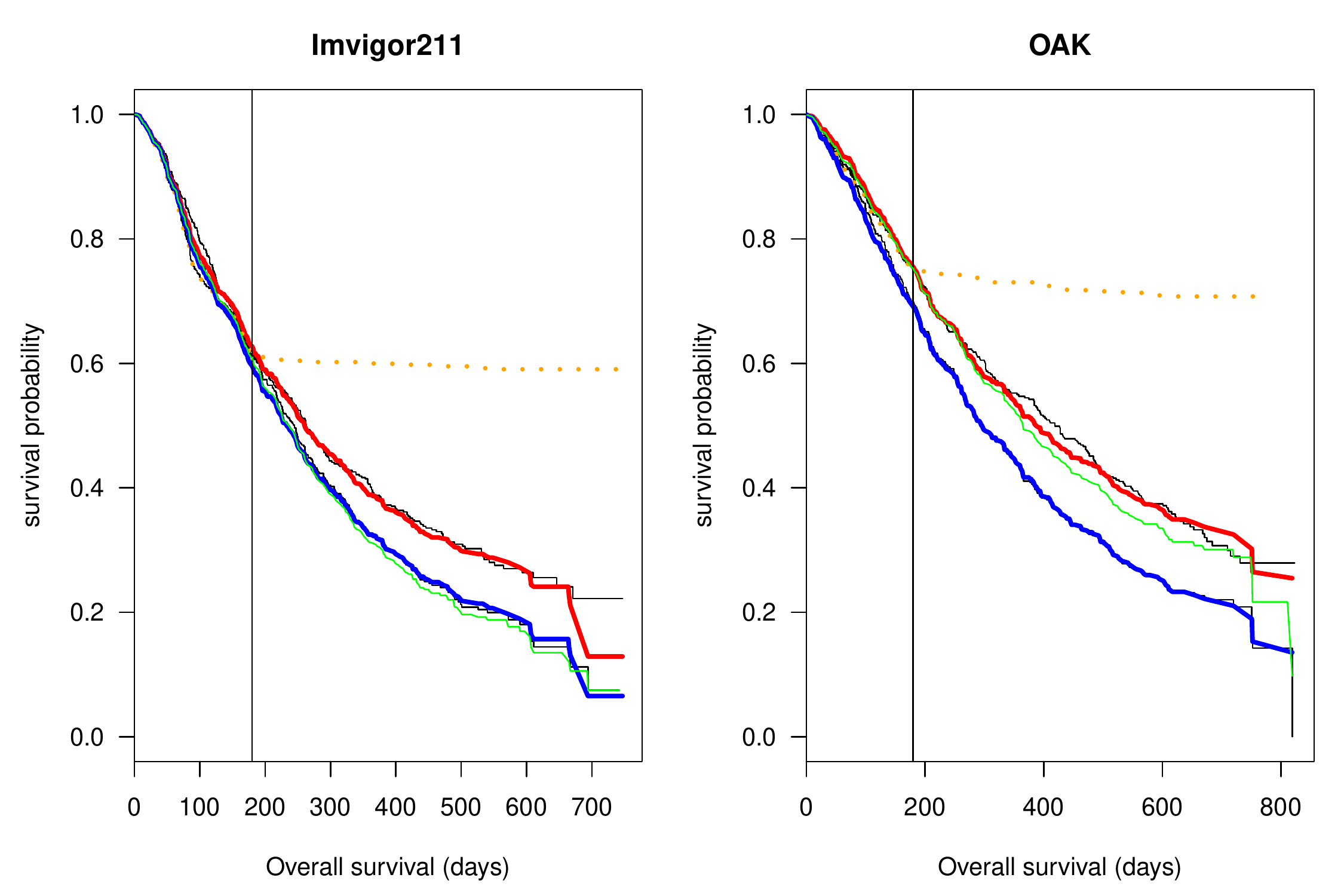}
\label{OAKIMCensoredfig}
\end{center}
\end{figure}

Figure \ref{OAKIMCensoredfig} shows the following estimates of the OS survival function: The black line depicts the KM estimate for the full dataset. The orange line is the KM estimate for OS of the experimental arm, where all post-PD events after 180 days are censored at day 180 or at one day after progression, whichever comes last. The green line is the OS prediction from a PH model, where deaths in the experimental arm are censored if beyond 180 days after randomization. In this model, the baseline hazard was estimated from both experimental and control arm, and the HR between the arms was used to derive the experimental OS survival function estimate. Finally, the red line is the OS survival function prediction based on the multistate model, taking all intermediate states into account. Especially in the Imvigor211 example, the advantage of the multistate model becomes obvious. Only deaths before the separation of the survival functions were available in the experimental arm, resulting in estimates of the experimental OS survival function (green line) close to the survival function of the control arm.

\subsection{Operating Characteristics for Oak based on 1000 simulated early phase trials of 40 patients}

After illustrating that the proposed method allows to predict the OS survival function in the experimental arm from our case studies, where the survival follow-up in the experimental arm is limited but the post-progression hazards are different, we illustrate that the multistate model is able to provide reliable OS estimates based on a dataset with limited sample size and limited follow up in the experimental arm. Similarly to the Cleopatra case study in Section \ref{sec:haz_same}, we simulate data and compute the operating characteristics for the Oak case study. Results for Imvigor211 are deferred to the appendix.

The total trial duration was assumed to be 1 year, including a recruitment period of half a year.

\begin{figure}[!ht]
\begin{center}
\caption{Multistate estimation of the OS survival function from 1000 simulated early phase trials of 40 patients (red lines) sampled from the experimental arm (left) and from the control arm (right) and using the complete control arm from Oak trial. Black lines represents the KM estimate of OS based on the full experimental (solid) and control (dashed) arms.}
\setkeys{Gin}{width=1\textwidth}
\includegraphics{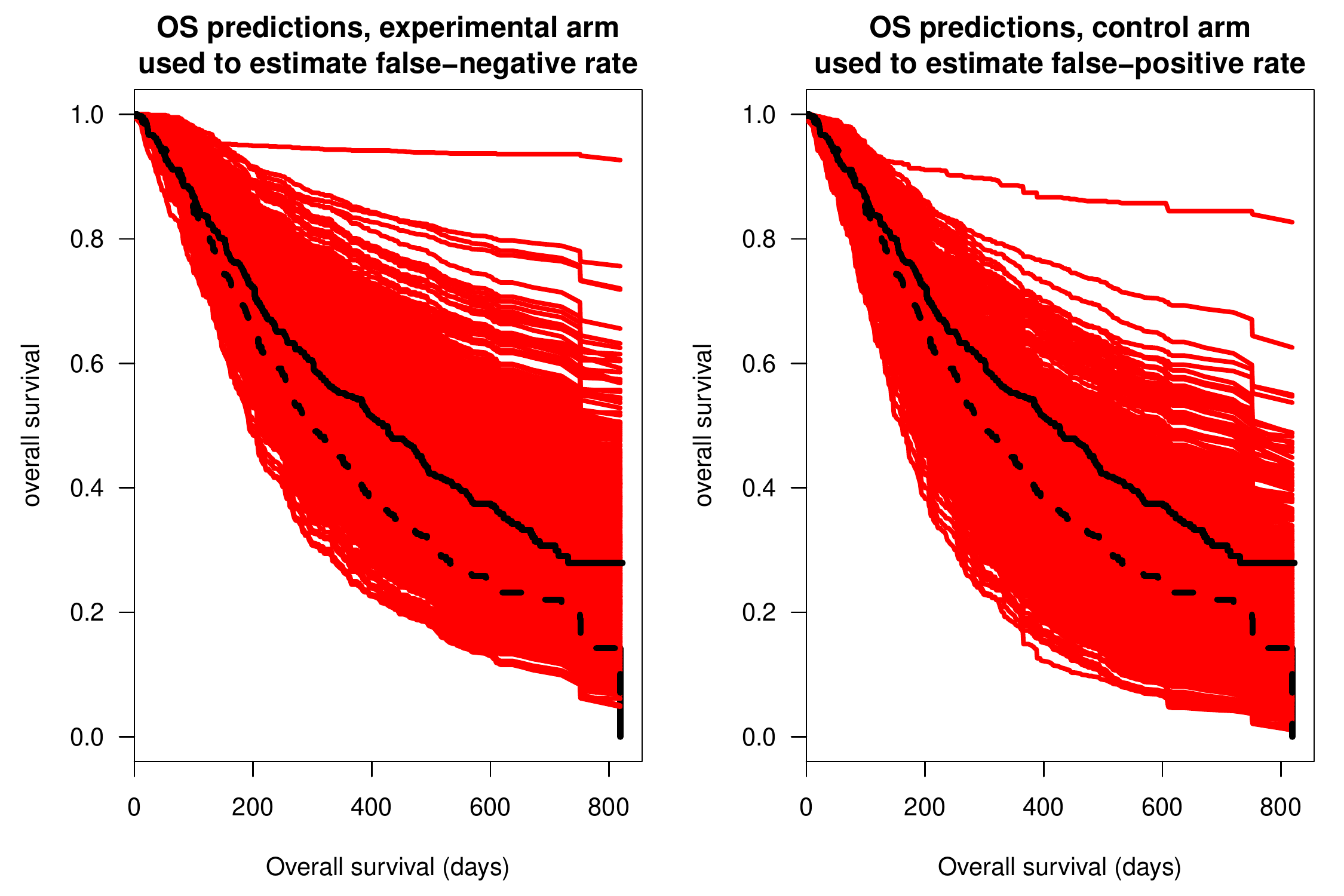}
\label{fig:OakSims}
\end{center}
\end{figure}

Figure \ref{fig:OakSims} and Table \ref{tab:OakSims} give the results. Despite a comparable response proportion in both arms, simulation results show that the prediction and decision making based on a multistate model based on limited sample size and follow up time is feasible and improves operational characteristics compared to decision making solely based on response proportion. Again, the average HRs over 1000 simulations are with values of 0.78 and 0.99, respectively, in line with expectations (see Table~\ref{tab:oak}).

Compared to the operating characteristics for Cleopatra in Table~\ref{CleopatraSIMtable} we find that the false-negative error is increased. This can be explained through the lower treatment effect, i.e. OS HR, in Oak.

\begin{figure}
\caption{table}{False-negative and false-positive rate of multistate decision rule based on 1000 simulated early phase trials of 40 patients sampled from the experimental (left) or control (right) arms in Oak.}
\label{tab:OakSims}
\begin{center}
\begin{tabular}{lcc}
\multicolumn{3}{c}{False-Negative rate} \\
 & HR target value &  \% HR $>$ target value  \\
\hline & $0.80$&0.44 \\
 & $0.85$&0.37 \\
 & $0.90$&0.30 \\
 & $1.00$&0.17 \\
 & $0.80$&0.44 \\ \hline\end{tabular}
\end{center}
\bigskip
\begin{center}
\begin{tabular}{lcc}
\multicolumn{3}{c}{False-positive rate} \\
 & HR target value &  \% HR $<$ target value  \\
\hline & $0.80$&0.24 \\
 & $0.85$&0.30 \\
 & $0.90$&0.38 \\
 & $1.00$&0.53 \\
 & $0.80$&0.24 \\ \hline\end{tabular}
\end{center}
\end{figure}

\section{Minimum required survival follow-up for the experimental arm}
\label{sec:minFU}

When designing a new trial and if we do not just borrow the post-progression hazard from the historical control data as in Section~\ref{sec:haz_same}, the length of follow up for death events needs to be determined to get a stable point estimate for the OS hazard ratio from the multistate model. In this section, we analyse this question via simulation.

At the time of design of the early phase trial, we expect only the full individual patient data for the reference control arm to be available. A multistate model can then be fitted to the control data and with this model, new data can be generated assuming potential differences in the transition hazards for the experimental arm compared to the control. The hypothesized HRs for each transition are to be determined based on the mechanism of action and its impact on the different transitions, and on the expected efficacy of the new compound.

Using the Oak control arm as an example, we simulate experimental arm data where no treatment effect is asssumed (i.e. HRs are set to 1) for all transitions, except for the ones from response to progression and progression to death.

This would reflect the experience with studies comparing CITs with chemotherapy, where a survival benefit was observed despite similar response proportions. In simulations, an identical assumed hazard from SD to response will lead to similar percentage of responders between both arms, the difference in the hazards from response to PD leads to a prolonged DOR, and with the additional effect assumed for the post-progression hazard, the difference in the OS survival functions can be tuned to become clinically meaningful.

The simulation study is then carried out as follows: First, the cumulative transition hazards are estimated in the control arm. Second, we simulate 1000 early phase trials based on these cumulative hazards and a hypothesized effect for each transition. For the hypothesis of an active treatment, the transition HRs were assumed to be 0.3 and 0.6 for the response to PD and post-progression transition, respectively, and 1 for all other transitions. This resulted in a hazard ratio between experimental and control, averaged over the 1000 simulations, of 0.726, a value close to the reported HR of the Oak trial. For the scenario of no treatment effect, all transition hazards are set to 1.
Third, we administratively censor the post-progression transition after a given follow up time. This censoring time is denoted {\it cut time} and chosen as 30, 60, 90, 120, 150, 180, 240, and 320 days after enrollment of the first patient. Only the post-progression transition is cut, because all patients are typically followed up for response and progression, as discussed in Section~\ref{subsec:model_ass}. For the patients that have transitioned to PD after the cut time, the post-progression transition is administratively censored at the PD time + 1 day. For the patients that have transitioned to PD before the cut time, but are censored or have an event after the cut time, the transition is administratively censored at the cut time. These steps mimic the typical restriction on the follow up of patients in early phase trials, where the follow up ends after progression. Fourth, we apply a new cut on the total trial duration. As median PFS in Oak and Imvigor211 was around 4 months, the analysis time was set to 6 month after last patient randomized, with recruitment assumed to be 1 year. This step is to mimic the situation in early development, where decisions about further development steps have to be taken after a reasonable amount of data (especially response and progression events) are accumulated.
Fifth, we estimate the multistate model on the modified simulated datasets and report the OS HR.
The sample size of the simulated early phase trial is again chosen to be 40.

\begin{figure}[!ht]
\begin{center}
\caption{Oak, Estimate and 95\% confidence interval for the OS HR as a function of the cut time (time from randomization of patient to censoring of transition 6). Left: Active treatment (Simulated HR = 0.3 and 0.6 for transitions 4 and 6, HR = 1 for other transitions). Right: No treatment effect (HR = 1 for all transitions).}
\setkeys{Gin}{width=1\textwidth}
\includegraphics{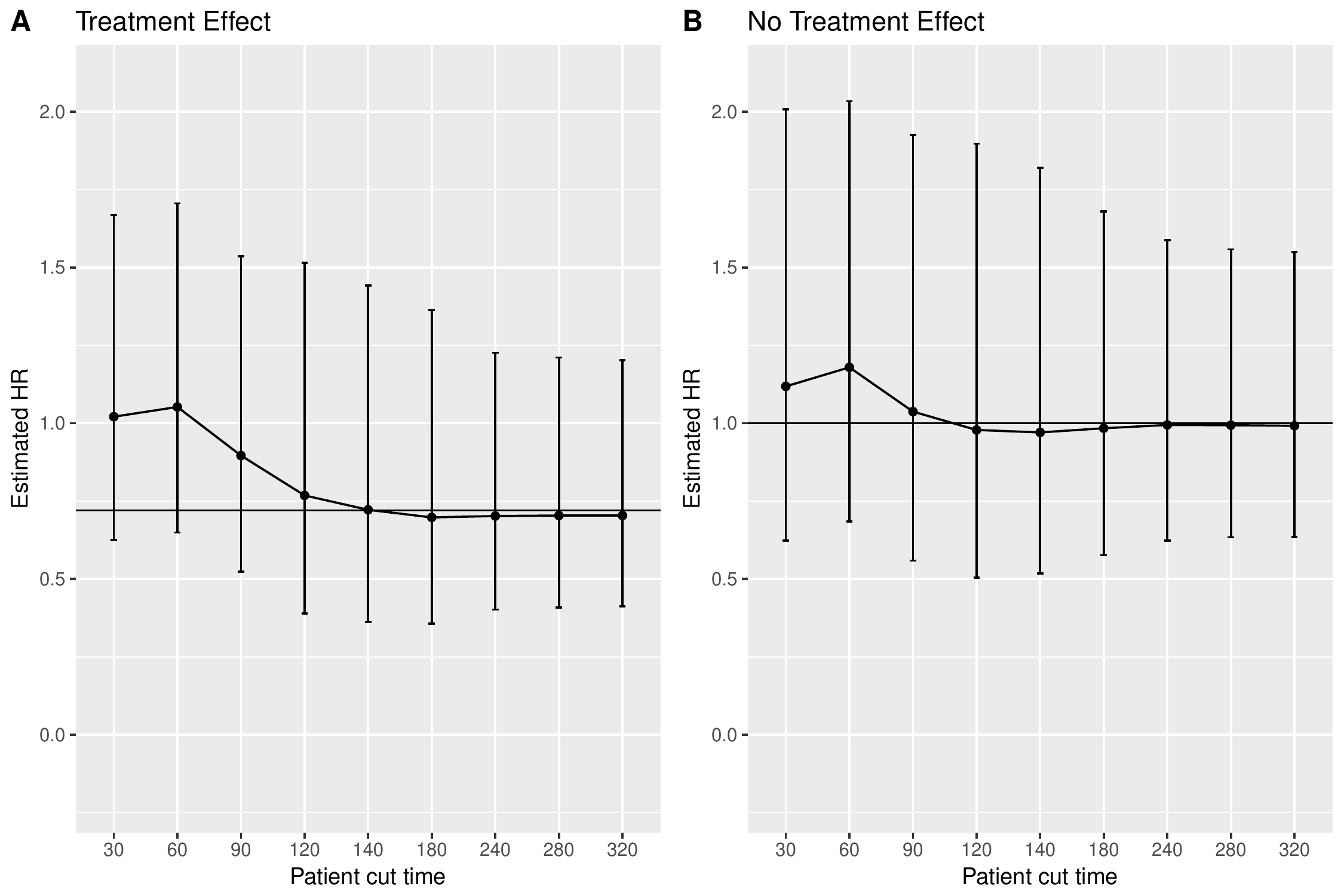}
\label{OAKSimCutTime}
\end{center}
\end{figure}

Figure \ref{OAKSimCutTime} reports the results. The average HR over the 1000 simulated early phase trials and a 95\% confidence interval are displayed as a function of the cut time. The left plot assumes an active experimental arm and the right plot assumes no treatment effect. We observe for both scenarios, i.e. when the experimental drug works as assumed above or does not work at all, that after a follow up of $\ge$180 days enough information on deaths would have been collected to get a sufficiently stable HR estimate. These results are comparable to a similar analysis done by \cite{huang_14}.

An alternative could be to wait until a given number of events is obtained on the transition from PD to death. However, this does not reflect the typical early phase trial situation, since these trials generally do not specify an endpoint of OS, making it hard to justify a clinical cutoff after a given number of death events. Typically, such trials fit in a development plan that requires a read-out at a fixed calendar time.

\section{Summary and Conclusion}
\label{summary}

A common way to evaluate the efficacy of a compound in an early phase trial in oncology is to evaluate tumor response and progression. With the emergence of new therapies using different mechanisms of action, this criteria may be less suitable to assess efficacy of a new therapy early. In addition, obtaining a reliable benchmark for the observed response proportion is complicated by changing treatment paradigms, e.g. of treating beyond progression. As a consequence, the link between the response proportion and the long-term Phase 3 endpoints (like OS) gets even weaker.

In this paper, we illustrate that decision-making after an early phase trial in oncology development can be considerably improved when not relying on differences in response rates only. In the Imvigor211 and Oak case studies, the molecule would not have moved forward based on a comparison of response rates.

Multistate models have been proposed earlier (see e.g. \citealp{ref:Broeet1999,ref:putter}) to analyse the efficacy of oncology treatments, and to gain a better understanding of the data. They are a natural way to take all transitions and transition times of a patient from trial entry to response, progression and/or death into account. Interestingly, in our experience, these standard oncology endpoints are often analysed quite independently of each other and not within a multistate framework. As we have shown, multistate models can be used to reliably predict survival differences, even in situations where the difference in response proportions is small. In addition, the measure provided by the multistate model, i.e. the OS HR, is more relevant to development programs and the patients than response, as it is directly related to the typically used primary endpoint of subsequent pivotal trials. Another advantage of multistate models is that it is straightforward to introduce transition-dependent covariates. A potentially relevant covariate could e.g. be change in tumor burden before progression as covariate for the post-progression transition. Exploring these generalizations is subject of ongoing research.

We explore scenarios where the post-progression hazard is either assumed to be identical between control and experimental arm, or proportional. Note that the post-progression, or $3 \to 4$, is only one of six transitions in our model described in Figure~\ref{fig:MM}. It is not too difficult to construct scenarios where all transitions have proportional hazards, but the resulting endpoint OS {\it does not}. In other words, assuming PH for all transition probabilities may be a reasonable assumption, as it may still give rise to a broad spectrum of overall survival functions that do not necessarily need to have proportional hazards.

In all our analyses, we have assumed a sample size of 40 patients for the early phase trial. As shown, this provides good operating characteristics for Cleopatra, in line with the quite substantial effect size of a OS HR of 0.64. Operating characteristics were less impressive for Oak and Imvigor211, because of their lower effect sizes. In the latter cases, one might want to consider increasing the sample size of the early phase trial, and/or increase the follow-up, in order to get more data to inform the OS prediction.

For our proposed method to work, one needs to be able to borrow from the control arm. That borrowed data is ideally sampled from the same population as the early phase trial itself, it must provide patient-individual information on response, progression, and death, and it must have sufficient (survival) follow-up such that we can reliably estimate the post-progression hazard function. The focus of our paper was to put the general concept to test of using a multistate model to compute predictions of the OS survival function and corresponding operating characteristics for an early phase decision. To make sure the control and experimental arm were from the same population, we used both from RCTs and artificially cut back the experimental arm data. This implies that calendar time and/or selection bias should not be present.

When prospectively applying the described multistate models to define early phase gating criteria, the control data will likely come from a different source than the actually collected experimental data from the early phase trial. One then needs to extend the proposed methodology such that the control and experimental arm populations are made suitably comparable, e.g. through propensity score methods, see e.g. \cite{austin_11, austin_14}. These extensions are subject of ongoing research.

\section{Code and data}
All analyses are performed using the \pkg{mstate} \proglang{R} package (\citealp{R}, \citealp{putter2007tutorial}).

\noindent {\bf{Conflict of Interest}}

\noindent {\it{Cleopatra, Imvigor211, and Oak are clinical trials that were sponsored by F. Hoffmann-La Roche. All authors are employees of this company, and own stocks.}}

\begin{appendix}

\section*{Appendix}

\subsection*{A.1.\enspace Detailed results of case studies}

Tables~\ref{tab:cleopatra}, \ref{tab:imvigor211}, and \ref{tab:oak} detail the efficacy outcomes reported in the primary publications of the three case studies. Table~\ref{tab:Cleotrans}, \ref{tab:Imvigortrans}, and \ref{tab:Oaktrans} provide the number of events for each transition of the proposed multistate model.

\begin{table}[htb]
\begin{center}
\caption{Cleopatra results, taken from \cite{baselga_12} and \cite{swain2015pertuzumab}. Note that Table 2 in \cite{baselga_12} reports response proportions for patients with measurable disease, while here proportions for all patients in Cleopatra are given.}
\begin{tabular}{rllll}
\hline
  & Pertuzumab + Trastuzumab  &  Trastuzumab &  HR (95\% CI)  & \\  \hline \\
  \textbf{Survival} &     N=402       &  N=406     \\
 {Overall Survival} &  &  &  0.64 (0.47,0.88)& \\
 {Progression-free Survival} &  &  &  0.62 (0.51,0.75)& \\
 \textbf{ Response} &     N=343       &  N=336  \\
 {Objective Response} & 275 (80.2\%)  &  233 (69.3\%)  &  \\
 {Complete Response} & 19 (5.5\%)  &  14 (4.2\%)  &  \\
 {Partial Response} & 256 (74.6\%)  &  219 (65.2\%)  &  \\
  {Stable Disease} & 50 (14.6\%)  &  70 (20.8\%)  &  \\
 {Progressive Disease} & 13 (3.8\%)  &  28 (8.3\%)  &  \\
 {Not assessed/assessable} & 5 (1.5\%)  &  5 (1.5\%)  &  \\
  \textbf{ Duration of Response} &     N=275       &  N=233     \\
  {Median (months, 95\% CI)} & 20.2 (16.0,24.0)  &  12.5 (10.0-15.0)  &  \\  \hline
\end{tabular}
\label{tab:cleopatra}
\end{center}
\end{table}

\begin{table}[htb]
\begin{center}
\caption{Cleopatra: Multistate model transition table.}
\label{tab:Cleotrans}
\begin{tabular}{llrr}
\hline
&  Transition & Pertuzumab + Trastuzumab & Trastuzumab \\
\hline \\
 & From SD to Response & 290 & 257 \\
 & From SD to PD &  63 &  98 \\
 & From SD to Death &   9 &  13 \\
 & From Response to PD & 130 & 138 \\
 & From Response to Death &   3 &   9 \\
 & From PD to Death &  57 &  74 \\
\hline\end{tabular}
\end{center}
\end{table}

\begin{table}[htb]
\begin{center}
\caption{Imvigor211 results, taken from \cite{powles2017atezolizumab}.}
\label{tab:imvigor211}
\begin{tabular}{rllll}
\hline
  &   Atezolizumab  &  Chemotherapy &  HR (95\% CI)  & \\  \hline \\
  \textbf{Survival} &     N=467       &  N=464     \\
 {Overall Survival} &  &  &  0.85 (0.73,0.99)& \\
 {Progression-free Survival} &  &  & 1.1 (0.95,1.26)& \\
 \textbf{ Response} &     N=462      &  N=461  \\
 {Objective Response} & 62 (13.4\%)  &  62 (13.4\%)  &  \\
 {Complete Response} & 13 (3\%)  &  16 (3\%)  &  \\
 {Partial Response} & 46 (10\%)  &  46 (10\%)  &  \\
  {Stable Disease} & 92 (20\%)  &  162 (35\%)  &  \\
 {Progressive Disease} & 240 (52\%)  &  150 (32\%)  &  \\
 {Not assessed/assessable} & 68 (15\%)  &  87 (19\%)  &  \\
  \textbf{ Duration of Response} &     N=62       &  N=62     \\
  {Median (months, 95\% CI)} & 21.7 (13.0,21.7)  &  7.4 (6.1-10.3)  &  \\
\hline \\
\end{tabular}
\end{center}
\end{table}

\begin{table}[htb]
\begin{center}
\caption{Imvigor211: Multistate model transition table.}
\label{tab:Imvigortrans}
\begin{tabular}{llrr}
\hline
&  Transition & Atezolizumab & Chemotherapy \\
\hline \\
 & From SD to Response &  62 &  62 \\
 & From SD to PD & 310 & 270 \\
 & From SD to Death &  74 &  91 \\
 & From Response to PD &  21 &  41 \\
 & From Response to Death &   2 &   8 \\
 & From PD to Death & 248 & 251 \\
\hline\end{tabular}
\end{center}
\end{table}

\begin{table}[htb]
\begin{center}
\caption{Oak results, taken from \cite{rittmeyer2017atezolizumab}.}
\label{tab:oak}
\begin{tabular}{rllll}
\hline
 &   Atezolizumab  &  Chemotherapy &  HR (95\% CI)  & \\
 \hline \\
  \textbf{Survival} &     N=425       &  N=425     \\
 {Overall Survival} &  &  &  0.73 (0.62,0.87)& \\
 {Progression-free Survival} &  &  &  0.95 (0.82,1.10)& \\
 \textbf{ Response} &     N=425      &  N=425  \\
 {Objective Response} & 58 (14\%)  &  57 (13\%)  &  \\
 {Complete Response} & 6 (1\%)  &  1 (<1\%)  &  \\
 {Partial Response} & 52 (12\%)  &  56 (13\%)  &  \\
  {Stable Disease} & 150 (35\%)  &  177 (42\%)  &  \\
 {Progressive Disease} & 187 (44\%)  &  117 (28\%)  &  \\
 {Not assessed/assessable} & 30 (7\%)  &  74 (17\%)  &  \\
  \textbf{ Duration of Response} &     N=58       &  N=57     \\
  {Median (months, 95\% CI)} & 26.3 (10,NE)  &  6.2 (4.9-7.6)  &  \\
\hline \\
\end{tabular}
\end{center}
\end{table}

\begin{table}[htb]
\begin{center}
\caption{Oak: Multistate model transition table.}
\label{tab:Oaktrans}
\begin{tabular}{llrr}
\hline
&  Transition & Atezolizumab & Chemotherapy \\
\hline \\
 & From SD to Response &  58 &  57 \\
 & From SD to PD & 306 & 249 \\
 & From SD to Death &  44 &  74 \\
 & From Response to PD &  26 &  40 \\
 & From Response to Death &   2 &   7 \\
 & From PD to Death & 224 & 212 \\
\hline\end{tabular}
\end{center}
\end{table}

\clearpage

\subsection*{A.2.\enspace Derivation of transition probabilities in four-state multistate model}

In this section we derive the four transition probabilities $P_{1  \to 4}, P_{1  \to 3  \to 4}, P_{1  \to 2  \to 4},$ and $P_{1  \to 2  \to 3  \to 4}$ that are used in Section~\ref{subsec:model_ass} to define $S_{OS}$.

In theory, to derive these transition probabilities one could follow the steps described e.g. in Section~A.2.5. of \cite{aalen_08}. This would entail writing down the transition intensity matrix and then solve the Kolmogorov forward equations. However, the transition probabilities can also be derived in an informal and intuitive way, see e.g. the paragraph after Eq. (27) in \cite{putter2007tutorial}. As an example, to derive $P_{1 \to 4}$, i.e. the probability to directly transition from State 1 to State 4, we can write
\bean
P_{1  \to 4}(0,t)  &=& \int_0^t  P_{11}(0,u) \lambda_{14}(u) P_{44}(u,t) du. \label{eq:P14a}
\eean Here, $P_{11}(0,u)$ gives the probability to remain in State 1 from time 0 to $s$, then at $u$ the patient transitions to State 4 with intensity $\lambda_{14}(u)$, and finally has to remain in State 4 until time $t$. Since State 4 in our model is absorbing, the latter probability is equal to 1, so that we can simplify \eqref{eq:P14a} to
\bea
P_{1  \to 4}(0,t)  &=& \int_0^t \exp\Bl(-\Lambda_{12}(u) - \Lambda_{13}(u) - \Lambda_{14}(u)\Br) \lambda_{14}(u)  du. \label{eq:P14b}
\eea
Eq. \eqref{eq:P134}-\eqref{eq:P234} can be derived similarly.
\bean
P_{1  \to 3  \to 4}(0,t) &=& \int_0^t  P_{11}(0,u)  \lambda_{13}(u)  P_{34}(u,t) du \nonumber \\
&=& \int_0^t \exp\Bl(-\Lambda_{12}(u) - \Lambda_{13}(u) - \Lambda_{14}(u)\Br) \lambda_{13}(u) \times \nonumber\\
&& \hspace*{1cm}\Bl[1-\exp\Bl(-(\Lambda_{34}(t) - \Lambda_{34}(u))\Br)\Br] du,
\label{eq:P134}  \\
P_{1  \to 2  \to 4}(0,t) &=& \int_0^t  P_{11}(0,u)  \lambda_{12}(u) P_{2  \to 4}(u,t) du  \nonumber \\
 &=& \int_0^t \exp\Bl(-\Lambda_{12}(u) - \Lambda_{13}(u) - \Lambda_{14}(u)\Br) \lambda_{12}(u)  P_{2  \to 4}(u,t) du,\label{eq:P124}   \\
 P_{1  \to 2  \to 3  \to 4}(0,t) &=& \int_0^t  P_{11}(0,u)  \lambda_{12}(u) P_{2  \to 3  \to 4}(u,t) du  \nonumber \\
 &=& \int_0^t \exp\Bl(-\Lambda_{12}(u) - \Lambda_{13}(u) - \Lambda_{14}(u)\Br) \lambda_{12}(u)  P_{2  \to 3  \to 4}(u,t) du,\label{eq:P1234}
\eean
where
\bean
P_{11}(0,t) &=& \exp\Bl(- \int_0^t  \lambda_{12}(u) +\lambda_{13}(u) + \lambda_{14}(u)du\Br) , \label{eq:P11}\\
P_{2  \to 4}(u,t) &=& \int_u^t \exp\Bl[-\Bl(\Lambda_{23}(v)- \Lambda_{23}(u)\Br) - \Bl(\Lambda_{24}(v)- \Lambda_{24}(u)\Br) \Br] \lambda_{24}(v) dv, \label{eq:P24} \\
P_{2  \to 3  \to 4}(u,t) &=& \int_u^t \exp\Bl[-\Bl(\Lambda_{23}(v)- \Lambda_{23}(u)\Br) - \Bl(\Lambda_{24}(v)- \Lambda_{24}(u)\Br) \Br]\lambda_{23}(v) \times \nonumber\\
 &&\hspace*{1cm} \Bl[1- \exp\Bl(- (\Lambda_{34}(v)- \Lambda_{34}(u))\Br) \Br]  dv \label{eq:P234}
\eean
The overall transition probability $P_{14}(0,t)$ from the initial state at time 0 to the death state at time $t$ consists of  the contributions of all possible paths from the initial to the death state, namely directly ($1 \to 4$), via the progressive state ($1 \to 3 \to 4$), or via the response state. In the latter case, two further scenarios have to be considered, namely a direct transition into death after response ($1 \to 2 \to 4$) or via an intermediate progressive event ($1 \to 2  \to 3 \to 4$).

\subsection*{A.3.\enspace Operating characteristics of decision rule for the IMVigor211 trial.}

\begin{figure}[!ht]
\begin{center}
\caption{Multistate estimation of the OS survival function based on 1000 simulated early phase trials of 40 patients (red lines) sampled from the experimental arm (left) and from the control arm (right) and using the complete control arm from Imvigor211 trial. Black lines represents the KM estimate of OS based on the full experimental (solid) and control (dashed) arms.}
\setkeys{Gin}{width=1\textwidth}
\includegraphics{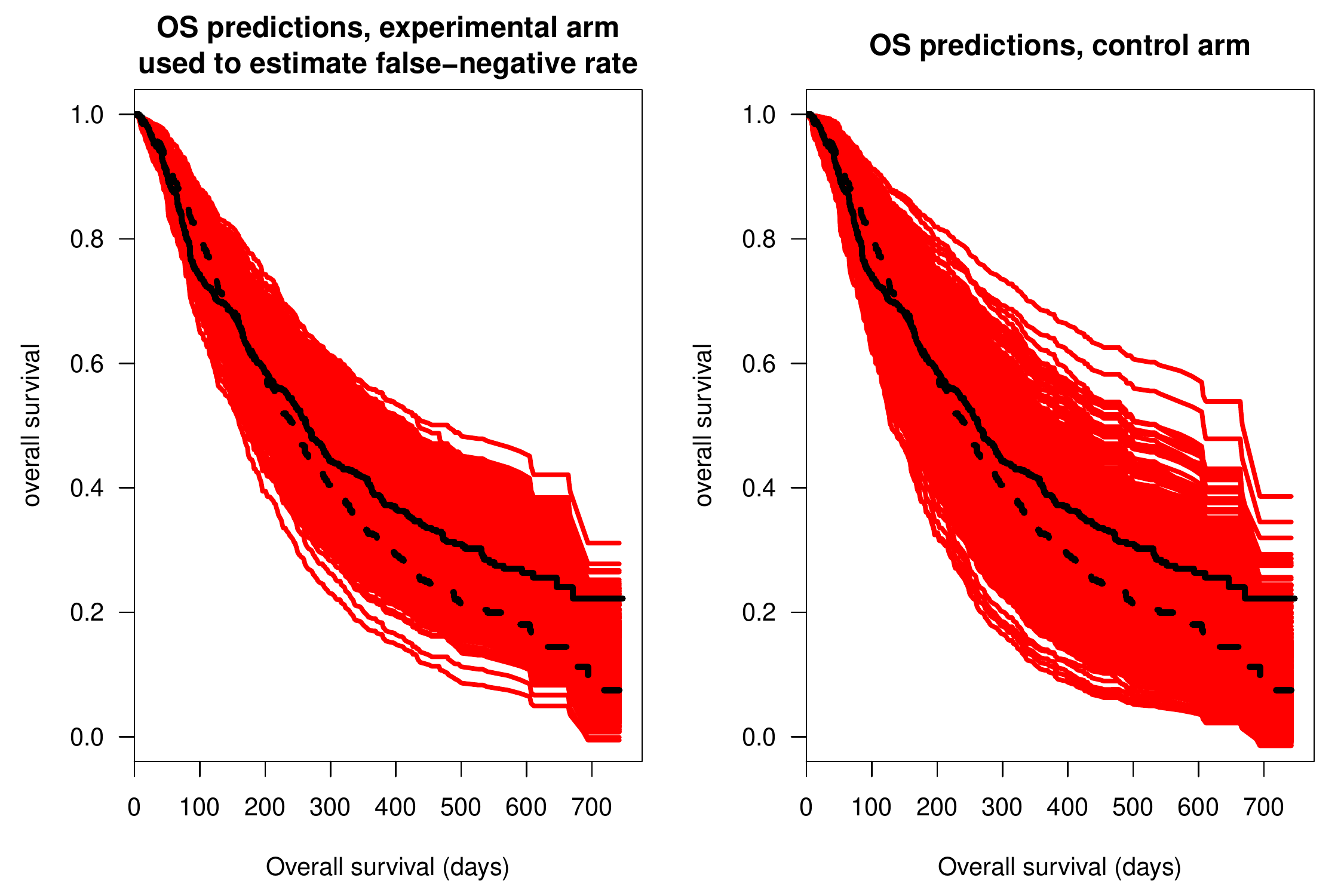}
\label{fig:IMvigorSims}
\end{center}
\end{figure}

In this section we present the operating characteristics of the decision rule based on simulations from the Imvigor211 dataset. We note, as detailed in Table~\ref{tab:imvigor211}, that the observed OS HR for this trial is 0.85, corresponding to a modest treatment effect. Therefore, it is not surprising that the operating characteristics are inferior to those seen in Cleopatra and Oak. Figure~\ref{fig:IMvigorSims} and Table~\ref{tab:IMvigorSims} show the predicted survival functions and the proportion of false-negative (left table) and false-positive (right table) decisions. The difference between the left and right table shows the sensitivity of the proposed decision rule in absence of a difference in OR proportion, and when the OS difference is small. These results further illustrate the usefulness of the multistate model for decision making. For completeness, the average HRs over 1000 simulations are 0.90 for the left and 0.99 for the right table, respectively.

\begin{figure}
\caption{table}{False-negative and false-positive rate of multistate decision rule based on 1000 simulated early phase trials of 40 patients sampled from the experimental (left) or control (right) arms and from the complete control in Imvigor211.}
\label{tab:IMvigorSims}
\begin{center}
\begin{tabular}{lcc}
\multicolumn{3}{c}{False-Negative rate} \\
 & HR target value &  \% HR $>$ target value  \\
\hline & $0.80$&0.76 \\
 & $0.85$&0.63 \\
 & $0.90$&0.50 \\
 & $1.00$&0.22 \\
 & $0.80$&0.76 \\ \hline\end{tabular}
\end{center}
\bigskip
\begin{center}
\begin{tabular}{lcc}
\multicolumn{3}{c}{False-positive rate} \\
 & HR target value &  \% HR $<$ target value  \\
\hline & $0.80$&0.19 \\
 & $0.85$&0.28 \\
 & $0.90$&0.37 \\
 & $1.00$&0.55 \\
 & $0.80$&0.19 \\ \hline\end{tabular}
\end{center}
\end{figure}

\end{appendix}

\bibliography{biblio}

\end{document}

%% file: commands.tex
\definecolor{light-gray}{gray}{0.85}

\newcommand{\blanco}[1]{ }

\renewcommand{\hat}{\widehat}

\newcommand{\bc}{\begin{center}}
\newcommand{\ec}{\end{center}}

\def\bea{\begin{eqnarray*}}
\def\eea{\end{eqnarray*}}
\def\be{\begin{equation}}
\def\ee{\end{equation}}
\def\bean{\begin{eqnarray}}
\def\eean{\end{eqnarray}}
\def\barr{\begin{array}}
\def\earr{\end{array}}
\def\bdes{\begin{description}}
\def\edes{\end{description}}
\def\bi{\begin{itemize}}
\def\ei{\end{itemize}}

\def\Bl{\Bigl}
\def\Br{\Bigr}

\def\ed{\end{document}}